\documentclass{aa}  
\usepackage{pifont}
\usepackage{graphicx}
\usepackage{txfonts}
\usepackage{lipsum}
\usepackage{natbib}
\usepackage{subcaption}        
\usepackage{lscape}             
\usepackage{placeins}          
\usepackage{xcolor}
\usepackage{hyperref}
\hypersetup{
    colorlinks=true,
    linkcolor=blue,
    citecolor=blue,
    filecolor=magenta,      
    urlcolor=cyan,
    }

\newcommand{\ZPEKTR}{\texttt{ZPEKTR}}
\defcitealias{TurisGallo2025}{Paper I}
\newcommand{\PaperI}{\citetalias{TurisGallo2025}}

\begin{document}

   \title{Extension and validation of the ZPEKTR code for determining the inclination angles of Be stars}
 
   \author{D. Turis-Gallo  \inst{1},
   M. Cur\'e \inst{1},  
   R. S. Levenhagen \inst{2}, C. Arcos \inst{1},  \and I. Araya \inst{3}}
   \institute{Instituto de F\'isica y Astronom\'ia, Facultad de Ciencias, Universidad de Valpara\'iso,  Av. Gran Bretaña 1111, Valpara\'iso, Chile.\\
           \email{daniela.turis@postgrado.uv.cl} \and
Departamento de Física, Universidade Federal de São Paulo, Rua Prof. Artur Riedel, 275, 09972-270, Diadema, SP, Brazil. \and
Centro Multidisciplinario de F\'{\i}sica,
Vicerrector\'{\i}a de Investigaci\'on,
Universidad Mayor, 8580745 Santiago, Chile}

   \date{Received 23 August 2026 / Accepted 21 October 2026}

  \abstract
    {Classical Be stars are non-supergiant B-type stars that rotate at a significant fraction of their critical velocity and are surrounded by an equatorial circumstellar gas disc. Their rapid rotation induces significant geometrical oblateness and gravity darkening. As a consequence, the emergent spectrum depends not only on the fundamental stellar parameters but also on the inclination angle.}
    {We aim to assess the reliability of the \texttt{ZPEKTR} code for determining inclination angles as a function of the stellar rotation rate, identify the minimum rotation rate required to obtain reliable results, and extend our previous analysis aimed at deriving consistent stellar parameters to a sample of 21 classical Be stars.}
    {We evaluated the reliability of the method using noise-injection tests on synthetic spectra spanning a broad range of rotational rates and inclination angles, with multiple statistically independent noise realisations at $S/N\sim100$. Next, we constructed a grid of gravity-darkened synthetic spectra using the \texttt{ZPEKTR} code, covering the atmospheric and rotational parameter space of classical Be stars. The stellar and rotational parameters were derived through the $\mathcal{Q}$ minimisation of the \ion{He}{I}\,$\lambda\lambda\, 4388, 4471, 6678$ \AA\ line profiles, after excluding lines affected by circumstellar disc contamination. The method was applied to 21 classical Be stars using archival BeSOS/PUCHEROS spectra and new observations from the LCO/NRES network.}
    {The rotation rate was reliably recovered across the tested inclinations, whereas the recovery of the inclination angle improved in line with an increasing rotation rate. Reliable inclination angle estimates were obtained for $V/V_{\rm c} \geq 0.46$. When applied to our sample of Be stars, the derived inclination angles reveal an excellent agreement with independent interferometric measurements, with $R^{2}=0.985$.}
    {The \texttt{ZPEKTR} code provides a promising spectroscopic method for constraining the inclination angles and rotational properties of rapidly rotating massive stars, offering a practical alternative to interferometric observations for larger samples.}

   \keywords{Stars: rotation -- Stars: emission-line, Be -- Stars: fundamental parameters -- Techniques: spectroscopic} 

   \titlerunning{Validation of the ZPEKTR code for determining inclination angles}\authorrunning{Turis-Gallo et al.}
   \maketitle
   
   \nolinenumbers

\section{Introduction} \label{sec:Intro}

B-type stars provide key insights into the evolution of intermediate- and high-mass stars, bridging the transition between A-type and O-type stellar populations. With effective temperatures of approximately 11\,000\,K–30\,000\,K and masses ranging from 2 to 16 M$_\odot$, they are key objects for studying radiatively driven winds, stellar rotation, and internal mixing processes (e.g. \citealt{Maeder2009}). Their intense ultraviolet flux and strong hydrogen and helium lines make them major contributors to the ionisation of the surrounding interstellar medium and reliable tracers of recent star formation in galaxies. A distinct subset of these objects is formed of classical Be stars, non-supergiant B-type stars that have exhibited Balmer lines in emission at least once \citep{Rivinius2013}. The emission lines originate from a gaseous, dust-free, Keplerian circumstellar disc formed by material ejected from the stellar surface. Rapid rotation is a key ingredient of the Be phenomenon, although additional mechanisms (e.g. non-radial pulsations, binarity, and localised mass-ejection events) are thought to help lift material into orbit, where its subsequent evolution is commonly described within the viscous decretion disc framework \citep{Rivinius2013, Ghoreyshi2023, Carciofi2025, LabadieBartz2025}. Classical Be stars are among the most rapidly rotating non-degenerate stars, often rotating close to their critical limit \citep{Townsends2004}. Their transient discs, spectroscopic and photometric variability, along with their polarimetry signatures, make them prime laboratories for studying angular momentum loss, mass-ejection mechanisms, and disc dynamics in massive stars \citep{CastanonEsteban2024, Pereira2024, LabadieBartz2025}.

The high rotational velocities of classical Be stars produce significant centrifugal deformation, causing their stellar surfaces to become oblate rather than spherical \citep{Rivinius2013, Levenhagen2024}. This deformation leads to variations in the local effective gravity across the stellar surface. This, in turn, produces gravity darkening: the poles become hotter and brighter than the equatorial regions owing to their higher effective gravity \citep{VonZeipel, Espinosa-Lara}. Consequently, the observed spectrum results from the integration of regions with different temperatures and surface gravities, making the emergent spectral features dependent not only on the stellar parameters and rotation rate, but also on the inclination angle of the rotation axis \citep{Fremat2005, Levenhagen2024}.

However, determining the inclination angles of rapidly rotating stars remains a challenge. Modern spectroscopic techniques can measure projected rotational velocities ($v\sin{i}$) down to a few tens of $km/s$ in early-B dwarfs \citep{simondiaz2014}. However, they are often limited by the degeneracy between the rotational velocity and inclination angle through the $v\sin i$, while interferometric observations, although providing direct constraints on stellar geometry, are currently available only for a limited number of bright nearby objects (e.g. \citealt{Cochetti2019}). Therefore, spectroscopic methods that incorporate rotational deformation and gravity-darkening effects are required to extend inclination determinations to larger samples of rapidly rotating stars.

In this study, we validate the \texttt{ZPEKTR} code \citep{Levenhagen2024} for determining the inclination angles and rotational parameters of classical Be stars. Our primary goal is to establish under which rotational conditions the inclination angle can be reliably recovered from gravity-darkened line profiles. We first investigate this through controlled noise-injection experiments on synthetic spectra. We then extend the analysis presented in \citet[ hereafter Paper I]{TurisGallo2025}, where the \texttt{ZPEKTR} code was applied to a small sample of only ten classical Be stars, by expanding the study to a sample of 21 classical Be stars with available spectroscopic observations and independent interferometric constraints. This expanded sample allowed us to further evaluate the performance of the method and its applicability for characterising the rotational properties of Be stars in general.

This two-step approach allows us to first quantify the intrinsic limitations of the method and subsequently test its performance on real observations. The remainder of this paper is organised as follows. Section~\ref{sec:data} describes the observational data and sample selection criteria. Section~\ref{sec:method} outlines the ZPEKTR code and the methods employed, including the $\mathcal{Q}$ fitting procedure and the noise injection tests. Section~\ref{sec:Results_Theo} presents the theoretical validation results. Section~\ref{sec:result_ZPEKTR} reports the derived parameters for the Be-star sample. A comparison with interferometric determinations and a broader discussion are provided in Section~\ref{sec:discussion}, followed by our conclusions in Section~\ref{sec:conclusion}.

\section{Data} \label{sec:data}

We selected our sample of classical Be stars from the work by \cite{Cochetti2019}, who reported circumstellar-disc inclination angles inferred from their spectro-interferometric modelling procedure. These values provide an external reference for the spectroscopic inclinations obtained with \texttt{ZPEKTR}. This yielded a total of 21 stars, 12 of which belong to the Be Star Spectroscopic Survey (BeSOS; \citealt{Arcos2018}), which is primarily comprised of classical Be stars. These observations were carried out between 2012 and 2015 using the Pontificia Universidad Católica High Echelle Resolution Optical Spectrograph (PUCHEROS), mounted on the 50\,cm telescope at the Santa Martina Observatory. Developed by the Center of AstroEngineering at the Pontificia Universidad Católica de Chile, PUCHEROS provides a resolving power of R $\simeq$ 17\,000 \citep{vanzi2012}.
%, reaching stars as faint as V=9 with a typical signal-to-noise ratio (S/N) of $\sim$70 .

The remaining nine stars were observed between October 2024 and January 2025 with the Las Cumbres Observatory Network of Robotic Echelle Spectrographs (NRES/LCO) mounted on 1\,m telescopes. These observations were carried out under the Chilean Telescope Allocation Committee (CNTAC) programme ID CLN2024B-008, delivering a resolving spectral power of R $\simeq$ 53\,000. The spectral extraction and reduction were performed using the open-source \textsc{Banzai-NRES} pipeline\footnote{\url{https://github.com/LCOGT/banzai-nres}}, which applies standard echelle reduction procedures including bias subtraction, flat-fielding, order tracing, wavelength calibration, and optimal extraction. The resulting spectra achieved an S/N of approximately 100 at 4450 $\AA$. The spectra used in this work are publicly available through their respective institutional archives: the reduced BeSOS spectra through the BeSOS web archive\footnote{\url{https://besos.ifa.uv.cl/}} \citep{Arcos2018}, and the raw and pipeline-reduced NRES spectra through the LCO Science Archive\footnote{\url{https://archive.lco.global/}} using the corresponding programme ID.

\section{Method} \label{sec:method}

\subsection{\texttt{ZPEKTR} code} \label{sec:Method:ZPEKTR}
To construct the gravity-darkened input model spectra, we employed the \textsc{ZPEKTR} code \citep{Levenhagen2024}. This code requires a grid of synthetic plane-parallel spectra covering the relevant range of atmospheric parameters as input. For stars with effective temperatures between 4\,000\,K and 10\,000\,K, we adopted LTE model atmospheres from \citet{Kurucz1979}. The corresponding spectral synthesis was carried out with the \textsc{SYNSPEC} code \citep{Hubeny1995ApJ...439..875H}. For stars in the intermediate temperature range, of 10\,000\,K to 15\,000\,K, we used \textsc{TLUSTY} to compute the atmospheric structures under the LTE approximation, explicitly accounting for the relevant atomic transitions in the calculation of the radiation field and level populations. For hotter stars with effective temperatures between 15\,000\,K and 33\,000\,K, we adopted the non-LTE \textsc{BSTAR2006} and \textsc{OSTAR2004} model atmosphere grids computed with \textsc{TLUSTY} \citep{Hubeny1988CoPhC..52..103H}. This approach ensures an adequate description of the atmospheric structure and emergent radiation field over the full temperature domain considered in this work.

In all cases, the adopted models spanned surface gravities in the range $2.0 < \log g < 5.0$ dex. The emergent flux distributions were then computed through spectral synthesis with \textsc{SYNSPEC} \citep{Hubeny1995ApJ...439..875H}, ensuring a homogeneous treatment of the final synthetic spectra across the different temperature regimes. The resulting spectral grid assumes solar metallicity throughout and comprises a total of $\sim$ 1.7 million synthetic spectra. The combination of the polar radius, equatorial radius, and mass values listed in Table~\ref{table:param_grid} yielded a rotation rate in the range $0.41\leq V/V_{\rm c} \leq 0.96$, spanning a broad interval of rotational rates inferred for classical Be stars and extending to the near-critical regime \citep[e.g.][]{Zorec2016, Rivinius2013}. The full parameter ranges adopted to construct the \texttt{ZPEKTR} model grid are summarised in Table~\ref{table:param_grid}. For the theoretical validation, we generated an additional set of models extending to lower rotation rates, as described in Sect. \ref{sec:Method:noise}.

\begin{table}[h!]
 \caption{Grid of \texttt{ZPEKTR} models.}                
 \label{table:param_grid}   
 \centering     
 \renewcommand{\arraystretch}{1.2} 
 \begin{tabular}{c|ccc }      
 \hline\hline               
 Parameter& Initial & Final & Step\\        
     \hline                     
    $T_{pole}$ [K] & 9000 & 33000 & 1000  \\    
    $i$ [°] & 2 & 90 & 2\\
    $R_{pole}$  $[R_{\odot}]$ & 2 &  15 & 1 \\
    $R_{eq}$  $[R_{\odot}]$ & 2.12 & 21.75 & irregular$^{*}$ \\
    M $[M_{\odot}]$ & 5 & 30 & irregular$^{*}$ \\
 
 \hline                                 
 \end{tabular}
 \begin{flushleft}
{\footnotesize $^{*}$ Non-uniform sampling}
 \end{flushleft}
 \end{table}
 
The code returns a comprehensive set of stellar parameters, including the mean effective temperature ($\langle T_{\rm{eff}} \rangle$), polar and equatorial effective temperatures ($T_{\rm{eff}}^{\rm{pole}}$; $T_{\rm{eff}}^{\rm{eq}}$), mean surface gravity ($\langle \log g \rangle$), polar and equatorial surface gravities ($\log g_{\rm{pole}}$; $\log g_{\rm{eq}}$), polar and equatorial radii ($R_{\rm{pole}}$; $R_{\rm{eq}}$), mass ($M$), projected rotational velocity ( $v\sin{i}$), equatorial and angular rotation rates relative to the critical value ($V/V_{\rm c}$; ($\Omega/\Omega_{c}$ ), and inclination angle ($i$).

\subsection{Spectral fitting}\label{sec:Method:chi2}

\begin{figure}
    \centering
    \includegraphics[width=0.8\linewidth]{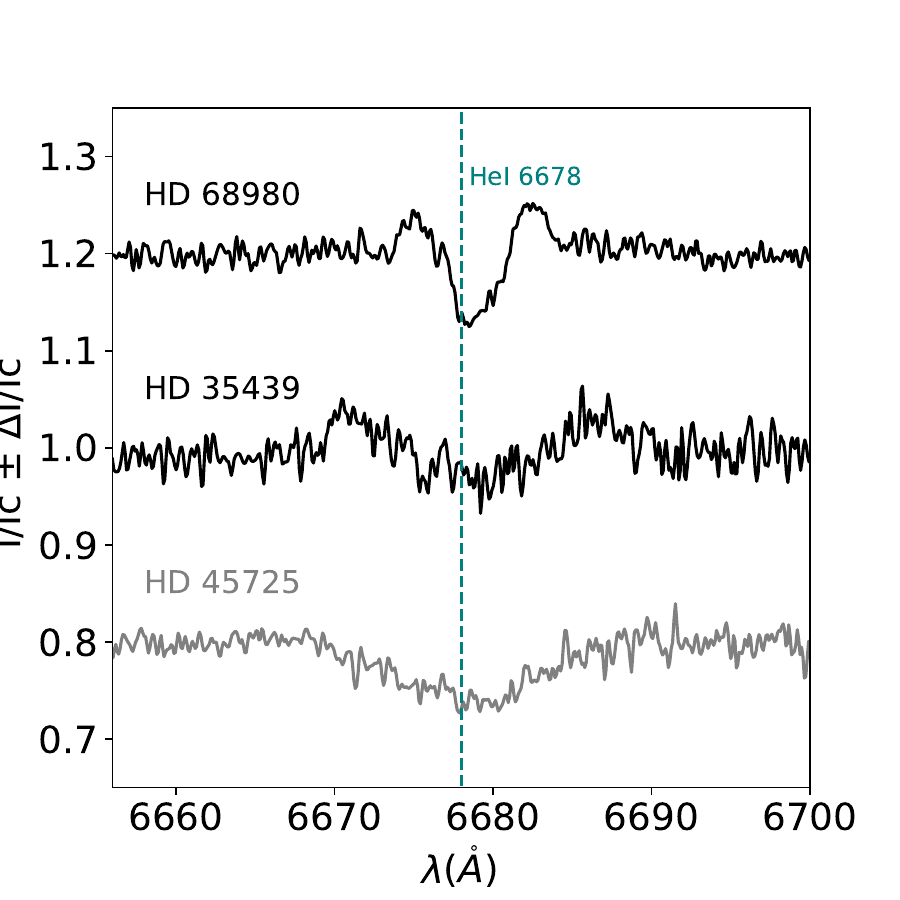}
    \caption{\ion{He}{I}\,$\lambda6678$ \AA\, profiles for three stars of the sample. The black spectra correspond to stars showing emission features, while the grey spectrum correspond to a purely absorption profile.}
    \label{fig:heI6678}
\end{figure}

To identify the best gravity-darkened model for each observed spectrum, we minimised the relative flux-weighted squared-residual merit function ($\mathcal{Q}$) across the grid of synthetic spectra generated with the \texttt{ZPEKTR} code. The fitting was performed over selected helium line profiles, specifically \ion{He}{I}\,$\lambda4388$ \AA, \ion{He}{I}\,$\lambda4471$ \AA, and \ion{He}{I}\,$\lambda6678$ \AA, which are sensitive to the stellar atmospheric parameters of B-type stars \citep{Solar2022}. These lines are primarily formed in the stellar photosphere and are therefore expected to be unaffected by the circumstellar disc, reflecting the photospheric conditions. However, Be stars are known to exhibit circumstellar emission that can still contaminate these photospheric lines through emission components or line infilling \citep[e.g.][]{Richardson2021, Balona2001}, as shown in Fig.~\ref{fig:heI6678} for the line \ion{He}{I}\,$\lambda6678$ \AA. Therefore, the number of spectral lines included in the fit was not fixed a priori. Instead, each line was visually inspected before the fitting procedure, and lines showing clear evidence of circumstellar contamination were excluded from the $\mathcal{Q}$ calculation. Depending on the characteristics of each stellar spectrum, the fit was performed using one, two, or all three available \ion{He}{I} lines, as appropriate for each case. The \ion{He}{I},$\lambda4471$ \AA\, line was always included in the fitting procedure, as this line generally does not present significant contamination from the circumstellar disc, following the methodology adopted in \PaperI. The specific set of spectral lines used in the fitting of each star is indicated in Table~\ref{tab:lines_obs}.

\begin{table}[h!]
\centering
\caption{\ion{He}{I} lines and observing programmes used.}

\label{tab:lines_obs}
\renewcommand{\arraystretch}{1.2}
\setlength{\tabcolsep}{3pt}
\begin{tabular}{l|cccc} \hline \hline

Star & \ion{He}{I},$\lambda4388$\AA\ & \ion{He}{I},$\lambda4471$\AA\ & \ion{He}{I},$\lambda6678$\AA\ & Obs. \\ \hline
HD 28497 & $\checkmark$ & $\checkmark$ & $\checkmark$ & \tablefootmark{2}\\
HD 30076 & $\checkmark$ & $\checkmark$ & $\checkmark$ &  \tablefootmark{2}  \\
HD 35439 & $\times$ & $\checkmark$ &  $\times$ & \tablefootmark{1} \\
HD 36576  & $\checkmark$ & $\checkmark$ & $\times$ & \tablefootmark{2}\\
HD 37490  & $\checkmark$ & $\checkmark$ & $\checkmark$ & \tablefootmark{2} \\
HD 37795  & $\checkmark$ & $\checkmark$ & $\checkmark$ & \tablefootmark{2} \\
HD 41335  & $\times$ & $\checkmark$ & $\times$ & \tablefootmark{1} \\
HD 45725  & $\times$ & $\checkmark$ & $\checkmark$ & \tablefootmark{1} \\
HD 50013  & $\checkmark$ & $\checkmark$ & $\times$ & \tablefootmark{2} \\
HD 60606  & $\checkmark$ & $\checkmark$ & $\checkmark$ & \tablefootmark{1} \\
HD 66194  & $\checkmark$ & $\checkmark$ & $\checkmark$ & \tablefootmark{2} \\
HD 68980  & $\checkmark$ & $\checkmark$ & $\times$ & \tablefootmark{1} \\
HD 89080  & $\times$ & $\checkmark$ & $\checkmark$ & \tablefootmark{2} \\
HD 91465  & $\times$ & $\checkmark$ & $\times$ & \tablefootmark{1} \\
HD 105435 & $\checkmark$ & $\checkmark$ & $\times$ &  \tablefootmark{2} \\
HD 120324 & $\checkmark$ & $\checkmark$ & $\times$ & \tablefootmark{1} \\
HD 158427 & $\checkmark$ & $\checkmark$ & $\times$ & \tablefootmark{1} \\
HD 209409 & $\times$ & $\checkmark$ & $\times$ & \tablefootmark{1} \\
HD 212076 & $\checkmark$ & $\checkmark$ & $\times$ & \tablefootmark{1} \\
HD 212571 & $\checkmark$ & $\checkmark$ & $\times$ & \tablefootmark{1}\\
HD 214748 & $\times$ & $\checkmark$ & $\times$ & \tablefootmark{1} \\ \hline
\end{tabular}
\tablefoot{ \\
\tablefoottext{1} {BeSOS} \\
\tablefoottext{2} {NRES/LCO} 

}
\end{table}

After selecting the suitable spectral lines for each star, we quantified the agreement between the observed and synthetic profiles through the merit function, $\mathcal{Q}$, computed as

\begin{equation}
    \mathcal{Q} = \sum_{j=1}^{N_\mathrm{lines}} \sum_{k=1}^{N_\lambda}
    \frac{\left[F_\mathrm{obs}(\lambda_k) - 
    F_\mathrm{mod}(\lambda_k)\right]^2}{ F_\mathrm{mod}(\lambda_k)},
    \label{eq:chi2}
\end{equation}

\noindent where $F_\mathrm{obs}(\lambda_k)$ and $F_\mathrm{mod}(\lambda_k)$ are the observed and synthetic normalised fluxes at a wavelength pixel, $\lambda_k$, then $N_\lambda$ is the number of wavelength points per line, and $N_\mathrm{lines} \in \{1, 2, 3\}$ is the number of uncontaminated lines available for a given star.

The model that minimised the total $\mathcal{Q}$ over all included lines was adopted as the best fit. Its associated parameters, such as effective temperature, surface gravity, rotation rate, and inclination angle, $i$, were taken as the derived stellar parameters. To estimate the uncertainties, we selected all the models within the first quartile of the $\mathcal{Q}$ distribution and computed the standard deviation of each parameter across this ensemble, following the methodology described in \PaperI. Since the spectra are continuum-normalised, the line profiles do not independently constrain the absolute stellar mass and radius. Therefore, to restrict the parameter space explored for each star, we adopted the stellar masses and radii reported by \cite{Cochetti2019, Cochetti2020} as reference constraints and considered ZPEKTR models within the corresponding parameter ranges. The masses and radii reported here should therefore be interpreted as properties of the best-fitting models within this constrained parameter space, rather than as independent measurements from the normalised line profiles alone.

We also assessed the effect of instrumental broadening on the different spectral resolutions present in our dataset by convolving the synthetic spectra with the corresponding instrumental profiles. No significant contribution was found in any of the cases examined, suggesting that instrumental broadening was negligible in the cases tested for stars with  $v\sin{i} \gtrsim 50$ km~s$^{-1}$, where rotational broadening dominates the line profile.

\subsection{Noise injection}\label{sec:Method:noise}

To evaluate the reliability and internal consistency of our method for determining stellar parameters, particularly the inclination angle according to the rotation rate, we performed a theoretical analysis based on synthetic spectra generated with the \texttt{ZPEKTR} code. The main goal of this test was to quantify the effect of observational noise on the recovery of stellar parameters and to establish the precision limits of the method as a function of S/N.

\begin{figure}[h!]
    \centering
    \includegraphics[width=0.95\linewidth]{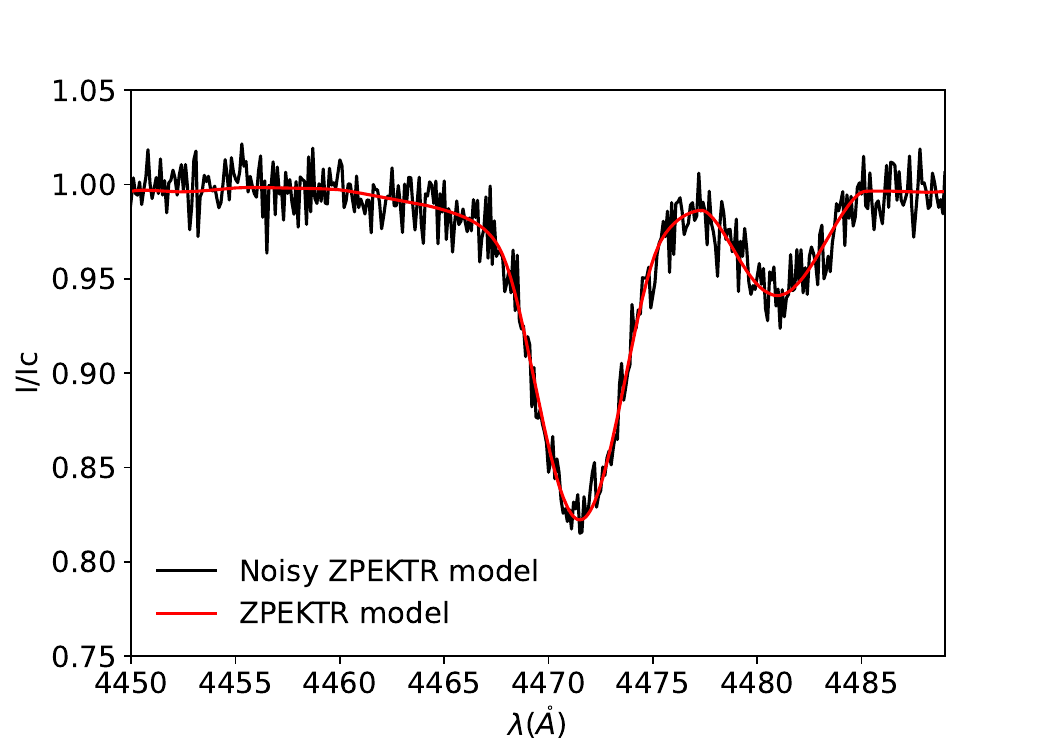}
    \caption{Example of a ZPEKTR synthetic spectrum with added Gaussian noise around the \ion{He}{I},$\lambda4471$ \AA. The original model (red line) is compared with the noisy realisation (black line) generated assuming S/N=100 ($\sigma$=0.01).}
    \label{fig:noisymodel}
\end{figure}

For this purpose, we generated an additional grid of $\sim$51\,300 synthetic spectra covering lower rotation rates than those considered in Sect.~\ref{sec:Method:ZPEKTR}. From this grid, we selected the models with $T_{\rm eff}=25\,000$ K, $R_{\rm pole}=7\,R_{\odot}$, and $M=9\,M_{\odot}$ to perform the noise analysis. For these fixed stellar parameters, the rotational configurations covered $V/V_{\rm c}=0.27$--$0.93$ and inclination angles between $15^{\circ}$ and $85^{\circ}$. The selected models therefore represent a range of rotational rates and viewing geometries within the new grid, while the full grid also includes models with different combinations of stellar parameters. To simulate the observational conditions, Gaussian noise was added to each synthetic spectrum according to 
\begin{equation}
    F_{\rm noisy}(\lambda) = F_{\rm model}(\lambda) + \mathcal{N}(0, \sigma),
\end{equation}
where $\mathcal{N}(0,\sigma)$ denotes a Gaussian distribution with a zero mean and standard deviation, $\sigma$. The noise level was determined from the adopted S/N given by $\sigma~=~<F_{\rm model}>/~{(S/N)}$, where $\langle F_{\rm model} \rangle$ corresponds to the mean continuum flux of the synthetic spectrum. Because the spectra are normalised, we have $\langle F_{\rm model}\rangle \approx 1$, which allows us to approximate the noise level as $\sigma \approx 1/(S/N)$. Therefore, for the adopted $S/N=100$, the corresponding noise level is $\sigma=0.01$. Fig.~\ref{fig:noisymodel} illustrates the effect of this noise addition on one of the synthetic spectra.

The resulting noisy spectra were analysed using the standard fitting procedure described in Sect.~\ref{sec:Method:chi2}, based on a $\mathcal{Q}$ minimisation using the \ion{He}{I},$\lambda4471$ \AA\, line, which is expected to be minimally affected by circumstellar disc contamination in the Be stars. For each synthetic spectrum, the best-fitting model was identified and the entire procedure was repeated 100 times to evaluate the statistical stability of the recovered parameters. This number of realisations was chosen to balance computational cost and the stability of the resulting uncertainties. A different random seed was generated for each realisation to ensure a statistically independent noise pattern.

\section{Results}
In this section, we present the results obtained from the theoretical validation tests and the analysis of the Be star sample. First, we evaluated the capability of the \texttt{ZPEKTR} code fitting procedure to recover rotational parameters from synthetic noisy spectra. We then applied the method to observed Be stars to derive their stellar and rotational parameters, with particular emphasis on the determination of inclination angles.

\subsection{Theoretical validation} \label{sec:Results_Theo}

The theoretical validation tests were performed using the noisy synthetic spectra described in Sect.~\ref{sec:Method:noise}. The primary objective of this study was to evaluate the robustness of the fitting procedure for obtaining the inclination angle of rapidly rotating massive stars such as Be stars. We first assessed the recovery of the rotational rate and then investigated how reliably the inclination angle could be recovered across different rotation regimes. In general, the results demonstrate that the \texttt{ZPEKTR} code can recover the input parameters with good accuracy over a broad range of rotational configurations. As expected, the precision of the recovered parameters improved with increasing S/N, whereas lower-quality spectra led to larger dispersions and stronger parameter degeneracies. Next, we present the results for $S/N \sim 100$.   

\subsubsection{Results for $V/V_{\rm c}$}

\begin{figure}[h!]
    \centering
    \includegraphics[width=0.97\linewidth]{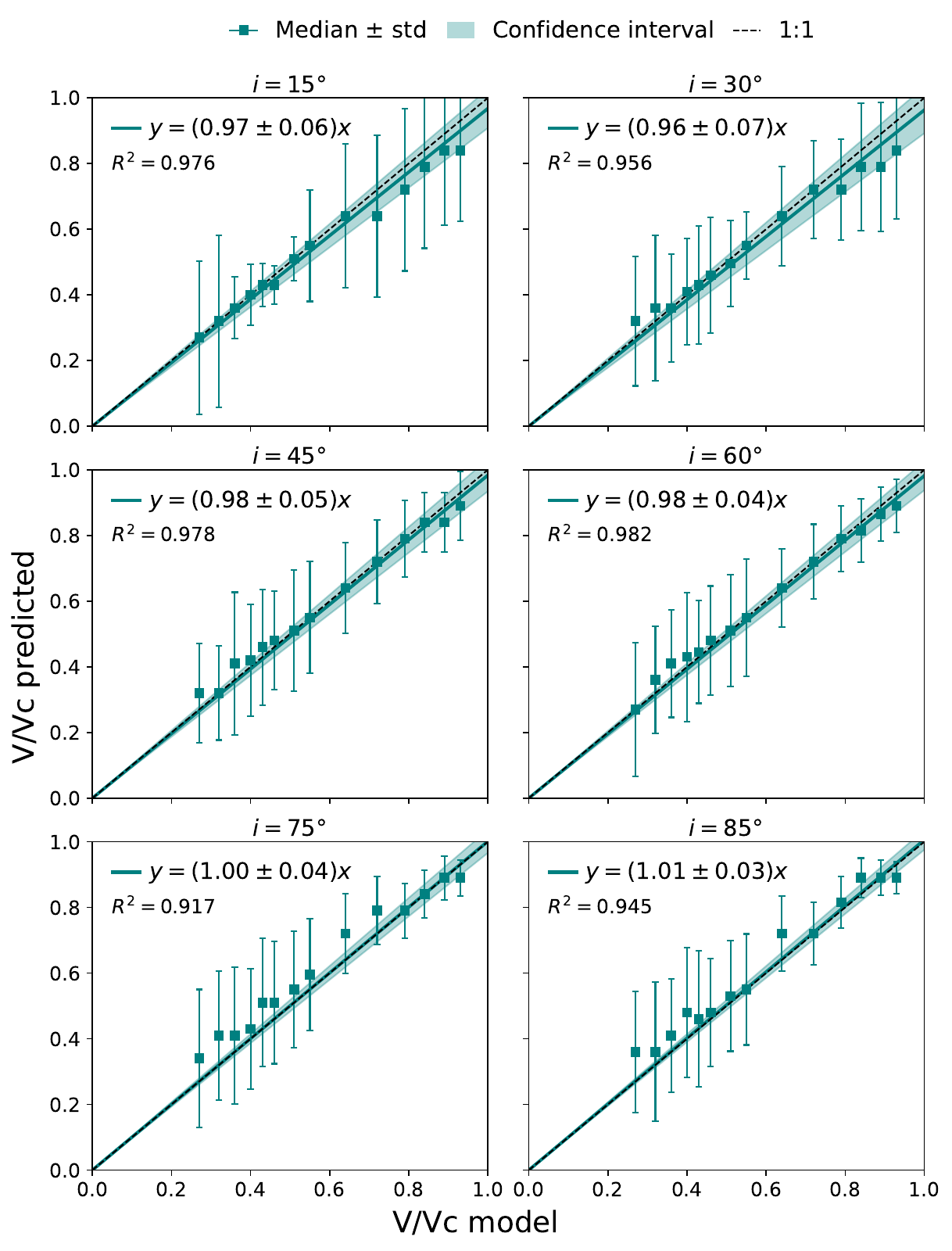}
    \caption{Modelled vs predicted $V/V_{\rm c}$  for inclination angles between 15$^{\circ}$ and $85^{\circ}$. Symbols indicate the median ± standard deviation. The solid line is a weighted linear fit expressed as $y = (a \pm \sigma)x$ and the shaded band is the fit confidence interval. The dashed line represents the 1:1 reference. $R^{2}$ values are shown per panel.}
    \label{fig:vvc_noise}
\end{figure}

Fig.~\ref{fig:vvc_noise} shows a comparison between the recovered and theoretical values of $V/V_{\rm c}$ for different inclination angles. Overall, the recovered rotational rates were in good agreement with the input models. To quantify this agreement, we performed linear fits between the original $V/V_{\rm c}$ values of the synthetic models and the recovered $V/V_{\rm c}$ values obtained from the fitting procedure. In all cases, the resulting correlations yielded coefficients of determination of $R^{2} > 0.9$, indicating a strong consistency.

\subsubsection{Inclination angles} \label{sec:result_i}
\begin{figure}[h!]
    \centering
    \includegraphics[width=0.98\linewidth]{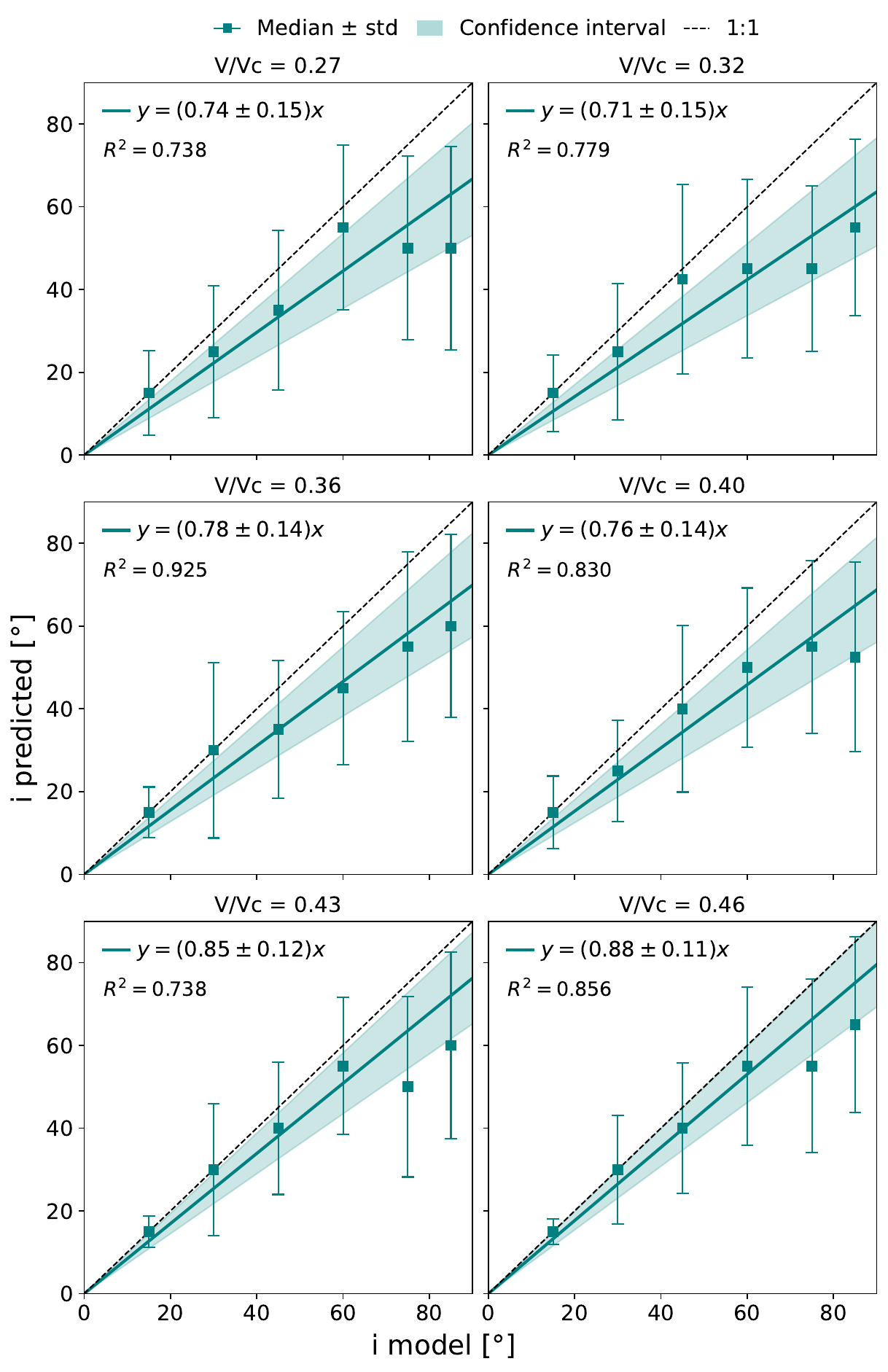}
    \caption{Comparison between the theoretical and recovered inclination angles for models with rotational rates between $V/V_{\rm c}=0.27$ and $0.46$. The symbols represent the median and associated standard deviation. The solid line shows the weighted linear fit and the shaded region indicates the confidence interval. The dashed line represents the 1:1 relationship between the model and recovered values. The corresponding coefficients of determination ($R^{2}$) are indicated in each panel.}
    \label{fig:i048}
\end{figure}

\begin{figure}[h!]
    \centering
    \includegraphics[width=0.98\linewidth]{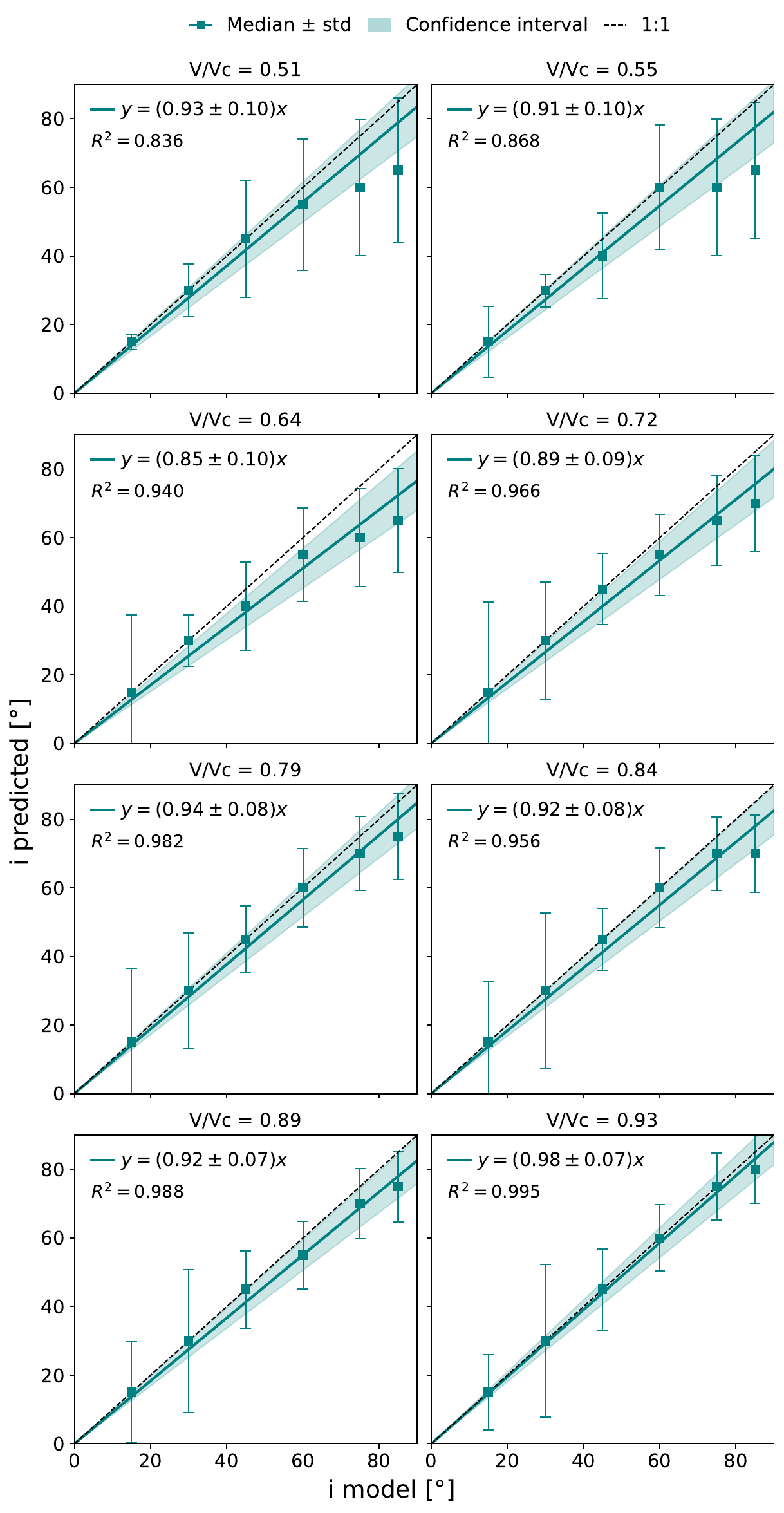}
    \caption{Same as Figure~\ref{fig:i048}, but for models with rotational rates between $V/V_{\rm c}=0.51$ and $0.93$.}
    \label{fig:i093}
\end{figure}

The recovery of the inclination angles exhibits a stronger dependence on the stellar rotation rate. Figs.~\ref{fig:i048} and ~\ref{fig:i093} compare the recovered inclination angles with the theoretical input values for different rotation rates.

The best agreement was obtained for rapidly rotating models, where the combined effects of gravity darkening and rotational deformation introduced more pronounced variations in the photospheric line profiles. Under these conditions, the inclination angle can be constrained with relatively small uncertainties, particularly for spectra with a high S/N.

However, at lower rotation rates, gravity-darkening and rotational deformation signatures become weaker, increasing the degeneracy between $V/V_{\rm c}$ and $i$. Consequently, different combinations of these parameters can produce similar line profiles, leading to broader distributions of recovered solutions. As the rotation rate increases, these effects become more pronounced, providing stronger spectroscopic constraints on the inclination angle and reducing the uncertainty in the recovered parameters.

Despite these limitations, the results indicate that the method can recover inclination angles with good overall consistency for intermediate and high rotational velocities. To assess the reliability of each linear fit, we computed the deviation of the fitted slope from unity in units of its uncertainty, $z = (1 - a) / \sigma_{a}$, where $a$ is the fitted slope and $\sigma_{a}$ is the associated standard error. We consider a fit acceptable when $z \le 1.5$ and $R^{2} > 0.8$, combining consistency with the ideal 1:1 relation and adequate explanation of the observed scatter. Table~\ref{tab:theo_z} summarises the fitted slopes, uncertainties, $R^{2}$ values, and z-scores for each $V/V_{\rm c}$. Of the 14 configurations analysed, 9 satisfied both criteria, corresponding to $V/V_{\rm c} \geq 0.46$. This trend suggests that the model recovers the inclination angle more reliably for stars rotating closer to the critical velocity.

\begin{table}[!ht]
\centering
\caption{Inclination angle fit parameters for each $V/V_c$.}
\label{tab:theo_z}
\centering        
\renewcommand{\arraystretch}{1.3}     
\begin{tabular}{cccccc} \hline \hline
$V/V_c$ &  $a$ & $\sigma_a$ & $R^2$ &  $z$ &     Accepted \\ \hline
   0.27 & 0.74 &       0.15 & 0.738 & 1.73 &     $\times$ \\
   0.32 & 0.71 &       0.15 & 0.779 & 1.93 &     $\times$ \\
   0.36 & 0.78 &       0.14 & 0.925 & 1.57 &     $\times$ \\
   0.40 & 0.76 &       0.14 & 0.830 & 1.71 &     $\times$ \\
   0.43 & 0.85 &       0.12 & 0.738 & 1.25 &     $\times$ \\
   0.46 & 0.88 &       0.11 & 0.856 & 1.09 & $\checkmark$ \\
   0.51 & 0.93 &       0.10 & 0.836 & 0.70 & $\checkmark$ \\
   0.55 & 0.91 &       0.10 & 0.868 & 0.90 & $\checkmark$ \\
   0.64 & 0.85 &       0.10 & 0.940 & 1.50 & $\checkmark$ \\
   0.72 & 0.89 &       0.09 & 0.966 & 1.22 & $\checkmark$ \\
   0.79 & 0.94 &       0.08 & 0.982 & 0.75 & $\checkmark$ \\
   0.84 & 0.92 &       0.08 & 0.956 & 1.00 & $\checkmark$ \\
   0.89 & 0.92 &       0.07 & 0.988 & 1.14 & $\checkmark$ \\
   0.93 & 0.98 &       0.07 & 0.995 & 0.29 & $\checkmark$ \\ \hline

\end{tabular}
\tablefoot{A fit is accepted when $z \leq 1.5$ and $R^2 > 0.8$.}
\end{table}

\subsection{Be stars} \label{sec:result_ZPEKTR}

\begin{figure*}[h!]
    \centering
    \includegraphics[width=0.9\linewidth]{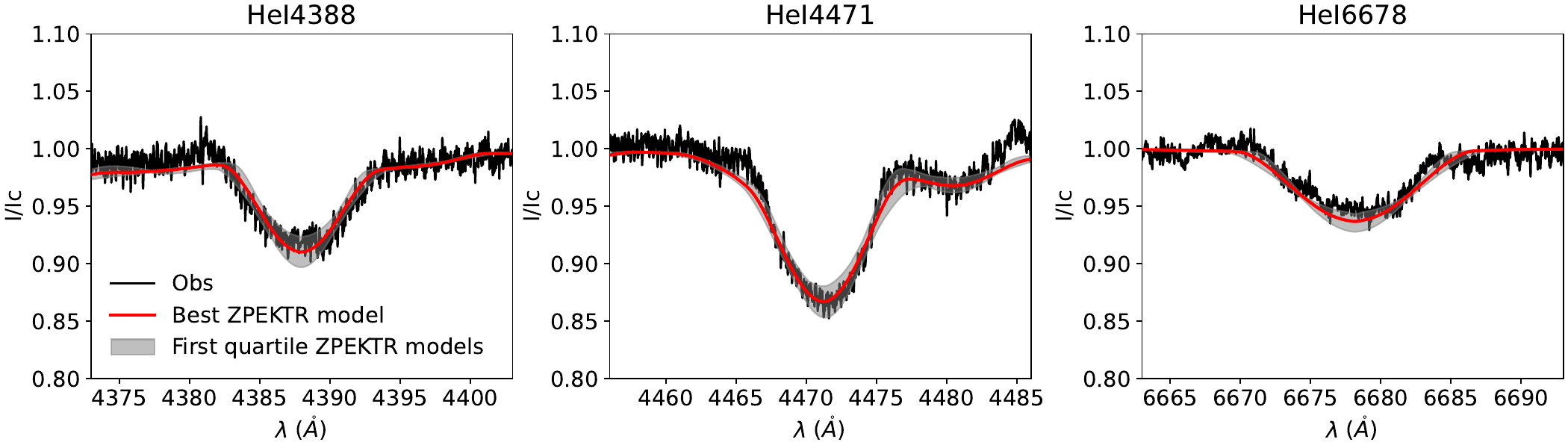}
    \caption{Spectral fitting results for HD 28497 using the \ion{He}{I},$\lambda4388$ \AA, \ion{He}{I},$\lambda4471$ \AA, and \ion{He}{I},$\lambda6678$ \AA\, line profiles. The black solid lines correspond to the observed spectra, while the red lines represent the best-fitting \texttt{ZPEKTR} model obtained through the $\mathcal{Q}$ minimisation procedure. The grey shaded region indicates the ensemble of models within the first quartile of the $\mathcal{Q}$ distribution, illustrating the uncertainties and degeneracies associated with the fitting process.}
    \label{fig:hd284997}
\end{figure*}

Having established the rotation regime in which the inclination angle can be reliably constrained, we went on to apply the \texttt{ZPEKTR} fitting procedure to the observed Be star sample described in Sect.~\ref{sec:data} to derive their stellar and rotational parameters. The analysis was carried out using the \ion{He}{I},$\lambda4388$ \AA, \ion{He}{I},$\lambda4471$ \AA, and \ion{He}{I},$\lambda6678$ \AA\, line profiles through the $\mathcal{Q}$ minimisation procedure described in Sect.~\ref{sec:Method:chi2}. In Fig.~\ref{fig:hd284997}, we give the spectral fits obtained for the three \ion{He}{I} lines for a representative star in the sample. The observed spectra were well reproduced by the best-fitting \texttt{ZPEKTR} model, while the ensemble of models within the first quartile of the $\mathcal{Q}$ distribution illustrated the parameter uncertainties and the range of acceptable solutions associated with the fitting procedure.

The derived stellar parameters of the 21 Be stars are listed in Tables.~\ref{tab:teff_logg} and ~\ref{tab:r_v_i}. Table~\ref{tab:teff_logg}   presents the effective temperature and surface gravity parameters affected by gravity darkening, including their polar, equatorial, and mean values. Table~\ref{tab:r_v_i} lists the corresponding stellar radii, projected rotational velocities, rotational rates, and inclination angles derived from the \texttt{ZPEKTR} fitting procedure. In general, the obtained parameters are consistent with previous spectroscopic studies of rapidly rotating B and Be stars, particularly those accounting for rotational deformation and gravity-darkening effects \citep[e.g.][]{Fremat2005, Arcos2018}, supporting the reliability of the adopted methodology and the gravity-darkened models. A detailed comparison between the inclination angles and $v\sin{i}$ values derived in this work and a previous determination is presented and discussed in Sect.~\ref{sec:discussion}. Overall, the inclusion of rotational deformation and gravity-darkening effects in the \texttt{ZPEKTR} models provides improved constraints on the rotational properties of Be stars and enables a more realistic interpretation of their photospheric line profiles.

\begin{table*}[h!]
\caption{Stellar parameters for the 21 Be stars obtained using the \texttt{ZPEKTR} models.}             
\label{tab:teff_logg}      
\centering        
\renewcommand{\arraystretch}{1.5}               
\begin{tabular}{l|ccccccc} \hline \hline
    Star & $\langle T_{\rm{eff}} \rangle$ & $T_{\rm{eff}}^{\rm{pole}}$ & $T_{\rm{eff}}^{\rm{eq}}$ & $\langle \log g \rangle$ & $\log g_{\rm{pole}}$ & $\log g_{\rm{eq}}$  \\
    & \small[K] & \small[K] & \small[K] & \small[dex] & \small[dex] & \small[dex] \\ \hline      % table heading
HD 28497 & 25839 $\pm$ 1664 & 28000 $\pm$ 1858 & 22410 $\pm$ 1984 & 3.97 $\pm$ 0.12 & 4.06 $\pm$ 0.12 & 3.83 $\pm$ 0.14 \\
HD 30076 & 18204 $\pm$ 1998 & 20000 $\pm$ 2204 & 16646 $\pm$ 1600 & 3.64 $\pm$ 0.12 & 3.74 $\pm$ 0.13 & 3.55 $\pm$ 0.14 \\
HD 35439 & 25460 $\pm$ 1450 & 27000 $\pm$ 1585 & 23319 $\pm$ 1764 & 3.86 $\pm$ 0.11 & 3.92 $\pm$ 0.11 & 3.77 $\pm$ 0.14 \\
HD 36576 & 22906 $\pm$ 1588 & 24000 $\pm$ 1740 & 20728 $\pm$ 1434 & 3.83 $\pm$ 0.16 & 3.89 $\pm$ 0.15 & 3.73 $\pm$ 0.17 \\
HD 37490 & 18175 $\pm$ 998 & 19000 $\pm$ 1070 & 17179 $\pm$ 922 & 3.48 $\pm$ 0.12 & 3.53 $\pm$ 0.12 & 3.42 $\pm$ 0.13 \\
HD 37795 & 12865 $\pm$ 350 & 14000 $\pm$ 429 & 10181 $\pm$ 674 & 3.45 $\pm$ 0.15 & 3.53 $\pm$ 0.16 & 3.23 $\pm$ 0.16 \\
HD 41335 & 24612 $\pm$ 1430 & 26000 $\pm$ 1617 & 21640 $\pm$ 1242 & 4.12 $\pm$ 0.09 & 4.18 $\pm$ 0.10 & 3.99 $\pm$ 0.09 \\
HD 45725 & 19909 $\pm$ 1249 & 21000 $\pm$ 1429 & 17478 $\pm$ 1116 & 3.87 $\pm$ 0.12 & 3.92 $\pm$ 0.11 & 3.73 $\pm$ 0.13 \\
HD 50013 & 22951 $\pm$ 1950 & 25000 $\pm$ 2359 & 20808 $\pm$ 1662 & 3.83 $\pm$ 0.17 & 3.92 $\pm$ 0.17 & 3.73 $\pm$ 0.17 \\
HD 60606 & 18092 $\pm$ 1508 & 19000 $\pm$ 1609 & 16409 $\pm$ 1261 & 3.83 $\pm$ 0.13 & 3.88 $\pm$ 0.14 & 3.72 $\pm$ 0.14 \\
HD 66194 & 18451 $\pm$ 1536 & 19000 $\pm$ 1950 & 17936 $\pm$ 1200 & 4.45 $\pm$ 0.19 & 4.48 $\pm$ 0.17 & 4.42 $\pm$ 0.21 \\
HD 68980 & 19909 $\pm$ 1407 & 22000 $\pm$ 1549 & 18311 $\pm$ 1496 & 3.35 $\pm$ 0.11 & 3.46 $\pm$ 0.12 & 3.26 $\pm$ 0.12 \\
HD 89080 & 12155 $\pm$ 669 & 13000 $\pm$ 716 & 9454 $\pm$ 521 & 3.82 $\pm$ 0.13 & 3.89 $\pm$ 0.13 & 3.59 $\pm$ 0.10 \\
HD 91465 & 23754 $\pm$ 1418 & 25000 $\pm$ 1502 & 20808 $\pm$ 2209 & 3.73 $\pm$ 0.10 & 3.79 $\pm$ 0.10 & 3.59 $\pm$ 0.13 \\
HD 105435 & 22951 $\pm$ 1151 & 25000 $\pm$ 1286 & 20808 $\pm$ 1173 & 3.96 $\pm$ 0.15 & 4.05 $\pm$ 0.17 & 3.86 $\pm$ 0.16 \\
HD 120324 & 21519 $\pm$ 1238 & 23000 $\pm$ 1369 & 20800 $\pm$ 1273 & 4.14 $\pm$ 0.19 & 4.22 $\pm$ 0.20 & 4.11 $\pm$ 0.21 \\
HD 158427 & 18512 $\pm$ 1338 & 20000 $\pm$ 1190 & 16646 $\pm$ 1253 & 4.13 $\pm$ 0.10 & 4.22 $\pm$ 0.11 & 4.02 $\pm$ 0.10 \\
HD 209409 & 12297 $\pm$ 1384 & 13000 $\pm$ 1483 & 10820 $\pm$ 1120 & 3.83 $\pm$ 0.30 & 3.89 $\pm$ 0.30 & 3.69 $\pm$ 0.28 \\
HD 212076 & 18554 $\pm$ 983 & 20000 $\pm$ 1032 & 17907 $\pm$ 983 & 3.96 $\pm$ 0.16 & 4.04 $\pm$ 0.16 & 3.92 $\pm$ 0.17 \\
HD 212571 & 27371 $\pm$ 1311 & 29000 $\pm$ 1371 & 26227 $\pm$ 1546 & 4.37 $\pm$ 0.19 & 4.44 $\pm$ 0.20 & 4.33 $\pm$ 0.22 \\
HD 214748 & 12233 $\pm$ 577 & 13000 $\pm$ 702 & 10820 $\pm$ 492 & 3.60 $\pm$ 0.17 & 3.66 $\pm$ 0.18 & 3.47 $\pm$ 0.17 \\

 \hline                                  
 \end{tabular}
 \end{table*}

\begin{table*}
\caption{Rotational parameters for the 21 Be stars obtained using the \texttt{ZPEKTR} models.}             
\label{tab:r_v_i}      
\centering        
\renewcommand{\arraystretch}{1.5}
\begin{tabular}{l|cccccccc} \hline \hline
    Star & $M$ & $R_{\rm{pole}}$ & $R_{\rm{eq}}$ & $v\sin{i}$ & $V/V_{\rm c}$ & $\Omega/\Omega_{c}$ & $i$  \\
    & \small[$M_{\odot}$] & \small[$R_{\odot}$] & \small[$R_{\odot}$] & \small[km/s] & & & [$^{\circ}$] \\ \hline
HD 28497 & 15.0 $\pm$ 2.8 & 6.00 $\pm$ 0.53 & 7.80 $\pm$ 0.83 & 359 $\pm$ 27 & 0.83 $\pm$ 0.07 & 0.96 $\pm$ 0.04 & 50 $\pm$ 8 \\
HD 30076 & 20.0 $\pm$ 3.8 & 10.00 $\pm$ 1.14 & 12.50 $\pm$ 1.39 & 218 $\pm$ 38 & 0.77 $\pm$ 0.07 & 0.93 $\pm$ 0.04 & 34 $\pm$ 13 \\
HD 35439 & 15.0 $\pm$ 3.5 & 7.00 $\pm$ 0.79 & 8.40 $\pm$ 0.99 & 299 $\pm$ 21 & 0.71 $\pm$ 0.04 & 0.88 $\pm$ 0.03 & 54 $\pm$ 11 \\
HD 36576 & 7.0 $\pm$ 1.9 & 5.00 $\pm$ 0.80 & 6.00 $\pm$ 1.05 & 280 $\pm$ 34 & 0.71 $\pm$ 0.08 & 0.88 $\pm$ 0.07 & 70 $\pm$ 12 \\
HD 37490 & 15.0 $\pm$ 4.5 & 11.00 $\pm$ 1.37 & 12.50 $\pm$ 1.46 & 202 $\pm$ 16 & 0.60 $\pm$ 0.07 & 0.79 $\pm$ 0.06 & 54 $\pm$ 11 \\
HD 37795 & 6.0 $\pm$ 1.4 & 7.00 $\pm$ 1.01 & 9.80 $\pm$ 1.38 & 247 $\pm$ 40 & 0.93 $\pm$ 0.06 & 0.99 $\pm$ 0.02 & 54 $\pm$ 13 \\
HD 41335 & 20.0 $\pm$ 2.2 & 6.00 $\pm$ 0.62 & 7.50 $\pm$ 0.71 & 468 $\pm$ 39 & 0.77 $\pm$ 0.04 & 0.93 $\pm$ 0.03 & 68 $\pm$ 12 \\
HD 45725 & 15.0 $\pm$ 1.5 & 7.00 $\pm$ 0.90 & 8.75 $\pm$ 1.24 & 380 $\pm$ 32 & 0.77 $\pm$ 0.04 & 0.93 $\pm$ 0.02 & 70 $\pm$ 13 \\
HD 50013 & 15.0 $\pm$ 4.8 & 7.00 $\pm$ 1.86 & 8.75 $\pm$ 2.23 & 260 $\pm$ 26 & 0.77 $\pm$ 0.04 & 0.93 $\pm$ 0.02 & 40 $\pm$ 7 \\
HD 60606 & 10.0 $\pm$ 1.3 & 6.00 $\pm$ 1.00 & 7.20 $\pm$ 1.14 & 297 $\pm$ 30 & 0.71 $\pm$ 0.04 & 0.88 $\pm$ 0.03 & 66 $\pm$ 14 \\
HD 66194 & 10.0 $\pm$ 2.1 & 3.00 $\pm$ 0.88 & 3.23 $\pm$ 1.07 & 231 $\pm$ 22 & 0.46 $\pm$ 0.14 & 0.64 $\pm$ 0.14 & 50 $\pm$ 11 \\
HD 68980 & 15.0 $\pm$ 3.6 & 12.00 $\pm$ 0.78 & 15.00 $\pm$ 0.91 & 154 $\pm$ 11 & 0.77 $\pm$ 0.04 & 0.93 $\pm$ 0.02 & 30 $\pm$ 5 \\
HD 89080 & 7.0 $\pm$ 0.8 & 5.00 $\pm$ 0.78 & 7.00 $\pm$ 0.92 & 362 $\pm$ 55 & 0.93 $\pm$ 0.07 & 0.99 $\pm$ 0.02 & 68 $\pm$ 9 \\
HD 91465 & 8.0 $\pm$ 1.1 & 6.00 $\pm$ 0.48 & 7.50 $\pm$ 0.87 & 307 $\pm$ 40 & 0.77 $\pm$ 0.11 & 0.93 $\pm$ 0.06 & 74 $\pm$ 11 \\
HD 105435 & 20.0 $\pm$ 5.7 & 7.00 $\pm$ 1.00 & 8.75 $\pm$ 1.24 & 300 $\pm$ 36 & 0.77 $\pm$ 0.04 & 0.93 $\pm$ 0.03 & 40 $\pm$ 12 \\
HD 120324 & 15.0 $\pm$ 6.4 & 5.00 $\pm$ 0.79 & 5.68 $\pm$ 0.90 & 151 $\pm$ 22 & 0.60 $\pm$ 0.02 & 0.79 $\pm$ 0.02 & 24 $\pm$ 9 \\
HD 158427 & 15.0 $\pm$ 4.1 & 5.00 $\pm$ 0.65 & 6.25 $\pm$ 0.87 & 344 $\pm$ 23 & 0.77 $\pm$ 0.04 & 0.93 $\pm$ 0.03 & 46 $\pm$ 11 \\
HD 209409 & 7.0 $\pm$ 1.8 & 5.00 $\pm$ 1.30 & 6.25 $\pm$ 1.51 & 299 $\pm$ 78 & 0.77 $\pm$ 0.04 & 0.93 $\pm$ 0.03 & 66 $\pm$ 12 \\
HD 212076 & 10.0 $\pm$ 3.4 & 5.00 $\pm$ 1.29 & 5.75 $\pm$ 1.53 & 108 $\pm$ 26 & 0.63 $\pm$ 0.04 & 0.82 $\pm$ 0.04 & 20 $\pm$ 6 \\
HD 212571 & 25.0 $\pm$ 8.0 & 5.00 $\pm$ 0.69 & 5.68 $\pm$ 0.86 & 281 $\pm$ 30 & 0.60 $\pm$ 0.07 & 0.79 $\pm$ 0.06 & 36 $\pm$ 9 \\
HD 214748 & 6.0 $\pm$ 1.3 & 6.00 $\pm$ 0.74 & 7.50 $\pm$ 0.82 & 239 $\pm$ 30 & 0.77 $\pm$ 0.04 & 0.93 $\pm$ 0.03 & 60 $\pm$ 13 \\
\hline
\end{tabular}
\end{table*}

\section{Discussion}
\label{sec:discussion}
The rotational rates derived from the \texttt{ZPEKTR} fitting procedure were robust across the full range of inclination angles considered in this study. As demonstrated by the noise injection tests presented in Sect.~\ref{sec:Results_Theo}, the recovered $V/V_{\rm c}$ values show strong consistency with the input models regardless of the viewing geometry, with coefficients of determination $R^{2}>0.9$ in all cases (Fig.~\ref{fig:vvc_noise}). This indicates that the $\mathcal{Q}$ minimisation procedure can constrain the rotation rate reliably, even in configurations where the inclination angle itself may be poorly determined, as is the case at low rotational velocities. The rotational rates listed in Table~\ref{tab:r_v_i} can therefore be considered a trustworthy estimate for all stars in the sample, independently of the uncertainties associated with the inclination angle determination. This result is particularly relevant for Be stars, for which an accurate knowledge of $V/V_{\rm c}$ is essential to characterise their proximity to the critical rotation limit and to interpret the physical mechanisms driving mass ejection and disc formation.

The results presented in Table~\ref{tab:theo_z} provide a quantitative estimate of the rotational regime in which the inclination angle can be reliably recovered. At low rotation rates, from $V/V_{\rm c}=0.27$ to 0.43, the fitted slopes remain systematically below unity and the corresponding z-scores exceed the adopted threshold, indicating a significant departure from the ideal 1:1 relation despite, in some cases, relatively high $R^{2}$ values. This behaviour reflects the increasing degeneracy between rotation rate and inclination when rotational deformation and gravity-darkening effects become less pronounced in the line profiles. At $V/V_{\rm c}=0.46$, the fit first satisfies both our acceptance criteria ($z \leq 1.5$ and $R^{2}>0.8$), with a slope of $0.88 \,(\pm 0.11$) and $R^{2}=0.856$. For higher rotation rates, the fitted slopes progressively approach unity and the correlation strengthens, reaching $R^{2}=0.982$ at $V/V_{\rm c}=0.79$ and $R^{2}=0.995$ at $V/V_{\rm c}=0.93$. These results indicate that the ability to recover the inclination angle is not governed by a sharp transition but, rather, that it improves progressively as the rotational effects on the stellar surface become more pronounced. We therefore identified $V/V_{\rm c}\simeq0.46$ as an approximate lower limit for reliable inclination recovery under the conditions tested here, while substantially more accurate constraints were obtained for stars rotating at $V/V_{\rm c}\gtrsim 0.79$.

\begin{figure}[h!]
    \centering
    \includegraphics[width=\linewidth]{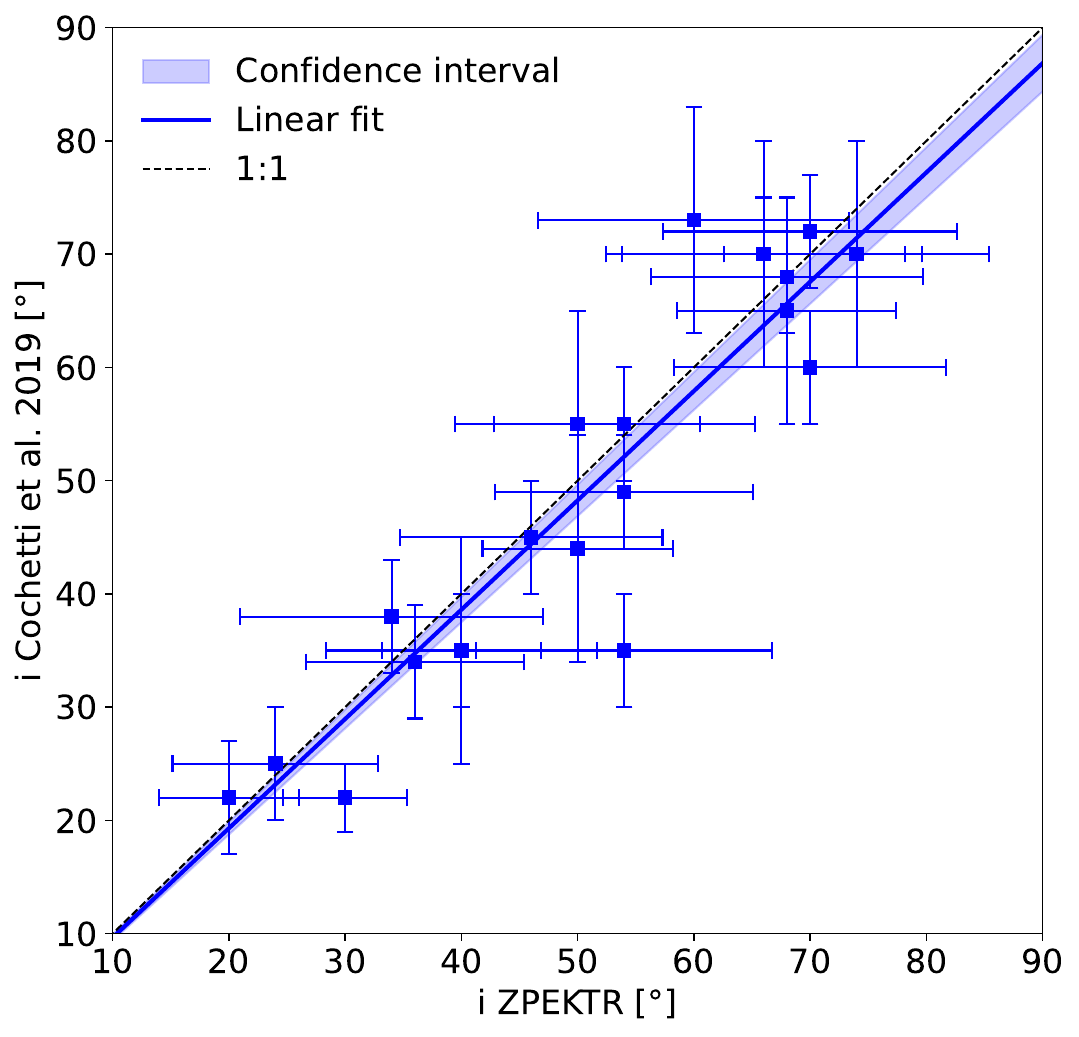}
       \caption{Comparison between the inclination angles derived from the \texttt{ZPEKTR} fitting procedure and the interferometric determinations of \cite{Cochetti2019} for the 21 stars in common. The dashed line represents the 1:1 relation, whereas the solid line shows the weighted linear fit. The shaded region indicates the confidence interval of fit. The close agreement between the two determinations yields ($R^{2}$=0.985)}
          \label{fig:fit_i_cochetti}
    \end{figure}

Fig.~\ref{fig:fit_i_cochetti} shows the comparison between the inclination angles derived in this work using the \texttt{ZPEKTR} fitting procedure and the independent interferometric determinations of \citet{Cochetti2019} for the 21 stars in common. The \texttt{ZPEKTR} inclination angles showed excellent agreement with the interferometric values across the full range sampled, from nearly pole-on to nearly equator-on configuration. A weighted linear fit to the data yields
\begin{equation}
y =  0.965\, {\small(\pm  0.028)\,} x, 
\end{equation}
with $R^{2} = 0.985$.

The slope is consistent with unity within $1.3\sigma$, suggesting that no significant systematic bias was introduced by the spectroscopic approach over the range of inclination angles probed by this sample. The slowest rotator in our sample is HD\,66194, for which we derived $V/V_{\rm c} = 0.46 \pm 0.14$ (Table~\ref{tab:r_v_i}). This places the star near the lower boundary of the rotational regime, where the inclination angle recovery becomes increasingly degenerate with the projected rotational velocity, as discussed in Sect.~\ref{sec:Results_Theo}. Nevertheless, the inclination angle derived from the \texttt{ZPEKTR} fitting procedure differs by only $5^{\circ}$ from the interferometric value of \citet{Cochetti2019}, a discrepancy that falls within the uncertainties of both determinations, suggesting that the method remains applicable even at the lower end of the sampled rotation rates.

We also compared our derived $v\sin{i}$ values with those reported in the literature by \cite{Glebocki2005}, \cite{Arcos2018}, and \cite{Solar2022}. The values of \cite{Glebocki2005} were estimated from empirical calibrations based on the full width at half maximum (FWHM) of the spectral lines. \cite{Arcos2018} derived $v\sin{i}$ by fitting the \ion{He}{I},$\lambda4471$ and \ion{Mg}{II},$\lambda4481$ line profiles with TLUSTY synthetic spectra broadened through rotational convolution, using a $\mathcal{Q}$ minimisation procedure, while simultaneously constraining $T_{eff}$ and $\log{g}$. \cite{Solar2022} derived $v\sin{i}$ using a Fourier-transform technique applied to several photospheric \ion{He}{I} lines, where the projected rotational velocity is determined from the first zero of the Fourier transform of the line profile. They considered different spectral windows corresponding to $2.0\sigma$, $2.5\sigma$, and $3.0\sigma$; here, we use their $2.0\sigma$ measurements for the comparison.

\begin{figure}[h!]
    \centering
    \includegraphics[width=\linewidth]{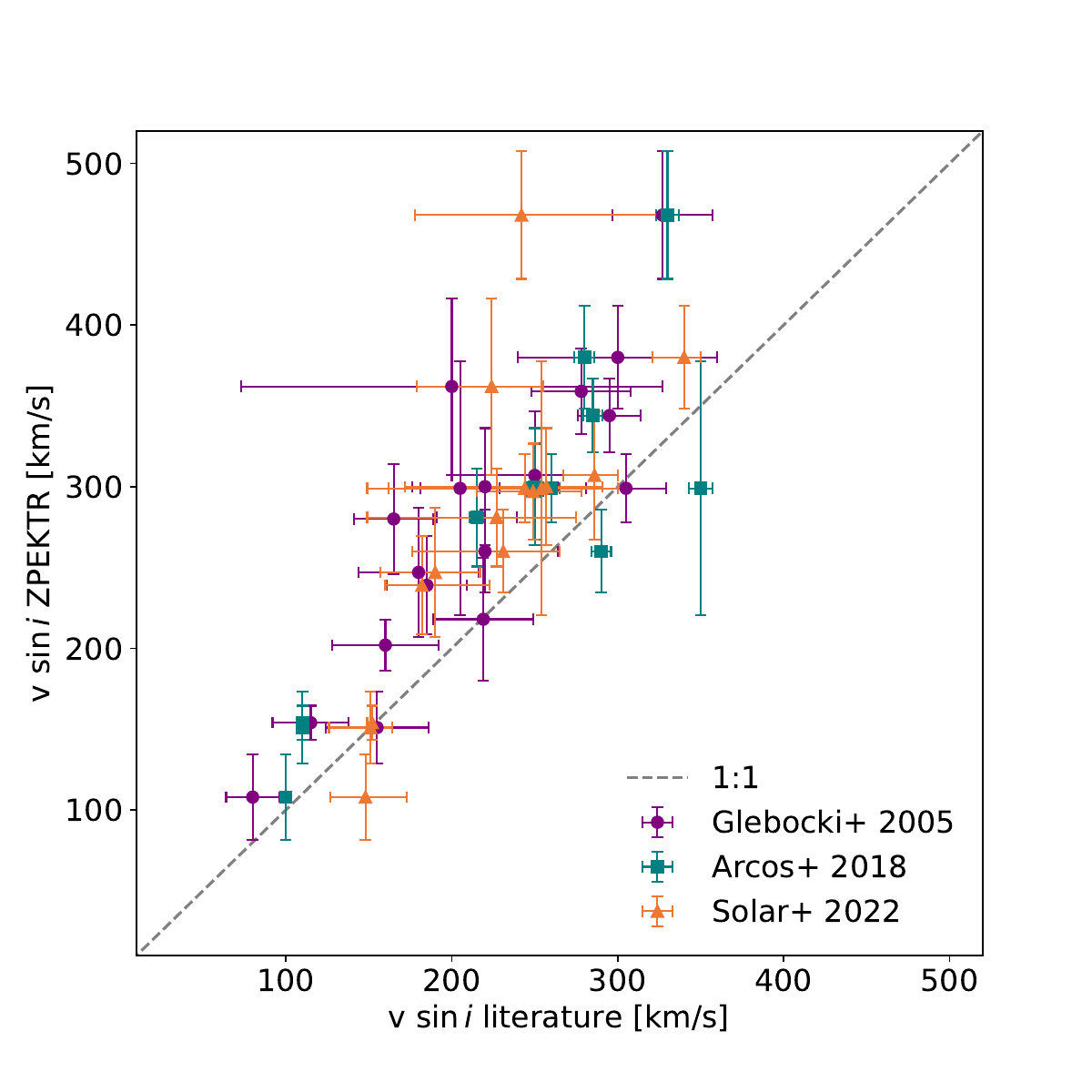}
       \caption{Comparison between the ($v\sin{i}$) values derived from the \texttt{ZPEKTR} fitting procedure and literature measurements from \cite{Glebocki2005}, \cite{Arcos2018} and \cite{Solar2022}. The dashed line represents a 1:1 relationship. The \texttt{ZPEKTR} values are, on average, 30\%, 20\%, and 21\% higher than the corresponding literature values, respectively.}
        \label{fig:vsini_literature}
    \end{figure}

These approaches assume a spherically symmetric stellar atmosphere and classical rotational broadening and therefore do not account for the effects of rapid rotation, such as gravity darkening and rotational deformation. In contrast, our method explicitly incorporates these effects into the spectral modelling. On average, our $v\sin{i}$ values are 30\% higher than those reported by \cite{Glebocki2005}, 20\% higher than those of \cite{Arcos2018}, and 21\% higher than those of \cite{Solar2022} (Fig.~\ref{fig:vsini_literature}).
This systematic difference is consistent with the expected impact of gravity darkening and rotational deformation on the helium line profiles, as shown by \cite{Levenhagen2014}. Neglecting these effects may therefore lead to systematically lower estimates of $v\sin{i}$ \citep{Stoeckley_1968, Townsends2004, Cranmer_2005}.

\section{Conclusion} \label{sec:conclusion}
In this study, we extended and validated the \texttt{ZPEKTR} code for the spectroscopic determination of inclination angles and rotational parameters in classical Be stars. Our main conclusions drawn are as follows:
\begin{enumerate}

\item We constructed a robust grid of $\sim$1.7 million gravity-darkened synthetic spectra with the \texttt{ZPEKTR} code, spanning polar temperatures from 9\,000 to 33\,000\,K, inclination angles from $2^{\circ}$ to $90^{\circ}$, and equatorial-to-critical velocity ratios that encompass the rotational regime characteristic of classical Be stars.
\item The theoretical validation through noise injection tests demonstrates that the rotational rates derived from the \texttt{ZPEKTR} fitting procedure are robust across the full range of inclination angles considered, with coefficients of a determination of $R^{2}>0.9$ in all cases. Therefore, the rotation rate can be reliably constrained independently of the inclination angle.

\item The recovery of inclination angles shows a strong dependence on the stellar rotation rate. Inclination angles can be robustly recovered only above the minimum rotation threshold. Reliable solutions were obtained for $V/V_{\rm c}\gtrsim0.46$, while the accuracy increased steadily with rotation, reaching $R^2 > 0.95$ for $V/V_{\rm c}\gtrsim0.79$.

\item The \texttt{ZPEKTR} fitting procedure was applied to a sample of 21 classical Be stars, combining archival BeSOS spectra and new LCO/NRES observations. The derived stellar and rotational parameters are consistent with those of previous spectroscopic studies reported in the literature, supporting the reliability of the gravity-darkened models.

\item A comparison of the derived inclination angles with the interferometric determinations of \citet{Cochetti2019} yields excellent agreement, with a weighted linear fit of $y = 0.965\,(\pm\,0.028)\,x$ and $R^{2} = 0.985$. The slope is consistent with unity within $1.3\sigma$, indicating that no statistically significant systematic offset is evident over the range of inclination angles probed by this sample.

\item Our estimations of $v\sin{i}$ are systematically higher than literature estimates, by 30\%, 20\%, and 21\% compared with \cite{Glebocki2005}, \cite{Arcos2018}, and \cite{Solar2022}, respectively, highlighting the importance of accounting for gravity darkening and rotational deformation when deriving projected rotational velocities of rapidly rotating stars.

\end{enumerate}

Taken together, these results establish the \texttt{ZPEKTR} fitting procedure as a reliable spectroscopic tool for determining the inclination angles and rotational properties of classical Be stars, providing a practical alternative for large stellar samples for which interferometric observations are not feasible. The excellent agreement with independent interferometric measurements demonstrates that gravity-darkened spectral modelling can recover inclination angles with high accuracy for rapidly rotating stars.

Despite the encouraging results presented above, this study is subject to several limitations. First, our validation was performed using synthetic spectra with $S/N \sim 100$ and solar-metallicity model atmospheres, which represent only a limited region of the parameter space relevant to Be stars. Second, the fitting was restricted to photospheric \ion{He}{I} lines largely unaffected by circumstellar contamination, thereby excluding the more strongly disc-affected features from the analysis. A systematic assessment of the performance of the method under different observing conditions, metallicities, and levels of disc emission therefore remains to be carried out and will be the subject of future work.

Future works will extend the analysis to a larger and more diverse sample of Be stars, investigate the applicability of the method across a broader range of spectral quality and metallicity, and assess its potential for other classes of rapidly rotating stars. In addition, a joint spectroscopic-photometric analysis will be pursued to evaluate the ability of the \ZPEKTR\, models to reproduce absolute fluxes, luminosities, and photometric observables, including the effects of inclination, extinction, and circumstellar emission. 

\begin{acknowledgements}

This work used the BeSOS catalogue, operated by the IFA-UV, Chile: http://besos.ifa.uv.cl, and is maintained by Fondecyt Nº 11190945. DT acknowledges support from \emph{ANID BECAS/DOCTORADO NACIONAL/2024-21240465}. MC  and CA acknowledge partial support from the Centro de Astrofísica de Valparaíso, Chile. MC, IA and CA thank the support from FONDECYT project N° 1230131 and N° 1261498. This study was made possible by using the AWS-U. Chile-NLHPC credits. Powered@NLHPC: This research was partially supported by the supercomputing infrastructure of NLHPC (ECM-02). This project has received funding from the European Union’s Framework Programme for Research and Innovation Horizon 2020 (2014-2020) under the Marie Skłodowska-Curie Grant Agreement No.~823734 and is also Co-funded by the European Union (Project 101183150 - OCEANS).
We thank the Chilean Telescope Time allocation for providing observations with the NRES spectrograph on the 1.0 m telescope at Las Cumbres Observatory (programme ID CLN2024B-008)
\end{acknowledgements}

\bibliographystyle{aa}
\bibliography{reference}

\end{document}